\documentclass[aps,reprint, prl, superscriptaddress, showpacs,nobibnotes]{revtex4-2}
\usepackage{graphicx}
\usepackage{hyperref}
\usepackage{bm,amsmath}
\usepackage{dcolumn}
\usepackage{natbib,xcolor}
\usepackage{gensymb}
\usepackage[version=4]{mhchem}

\begin{document}

\title{Elastoresistivity Signatures of Nematic Fluctuations in Layered Antiferromagnet \ce{CoTa3S6}}
\author{Tao Lu}
\affiliation{Division of Physics, Mathematics and Astronomy, California Institute of Technology, Pasadena, CA 91125, USA}
\affiliation{Institute of Quantum Information and Matter, California Institute of Technology, Pasadena, CA 91125, USA}
\author{Zili Feng}
\affiliation{Division of Physics, Mathematics and Astronomy, California Institute of Technology, Pasadena, CA 91125, USA}
\affiliation{Institute of Quantum Information and Matter, California Institute of Technology, Pasadena, CA 91125, USA}
\author{Mengxing Ye}
\affiliation{Department of Physics and Astronomy, University of Utah, Salt Lake City, UT 84112, USA}
\author{Takashi Kurumaji}
\affiliation{Division of Physics, Mathematics and Astronomy, California Institute of Technology, Pasadena, CA 91125, USA}
\affiliation{Institute of Quantum Information and Matter, California Institute of Technology, Pasadena, CA 91125, USA}
\author{Linda Ye}
 \email{lindaye@caltech.edu}
\affiliation{Division of Physics, Mathematics and Astronomy, California Institute of Technology, Pasadena, CA 91125, USA}
\affiliation{Institute of Quantum Information and Matter, California Institute of Technology, Pasadena, CA 91125, USA}

\date{\today}

\begin{abstract}

Nematic phases that break rotational symmetry are widely observed in quantum materials, and clarifying their origin and relationship with other symmetry-breaking phases remains an important but challenging task. In this work, we investigate nematic fluctuations in \ce{CoTa3S6} using elastoresistivity experiments to resolve the nature of the proposed nematic phase intertwined with collinear and non-coplanar antiferromagnetic orders. We observe a divergence-like antisymmetric elastoresistivity that rapidly develops below the stripe antiferromagnetic transition, consistent with a distinct nematic degree of freedom coupled to the magnetic order. 
While nematic fluctuations are strongly modulated by an external out-of-plane magnetic field and the onset temperature of resistivity anisotropy shows pronounced strain dependence, the antiferromagnetic transition temperatures remain nearly unchanged under either magnetic field or strain.
Additionally, complementary magnetoresistance measurements reveal characteristic signatures of three-state nematicity in a hexagonal system. 
Our findings demonstrate \ce{CoTa3S6} as a unique case of intertwined nematic and AFM orders with distinct origins. 

\end{abstract}

\maketitle

\begin{figure*}[t]
    \centering
    \includegraphics[width=0.9\textwidth]{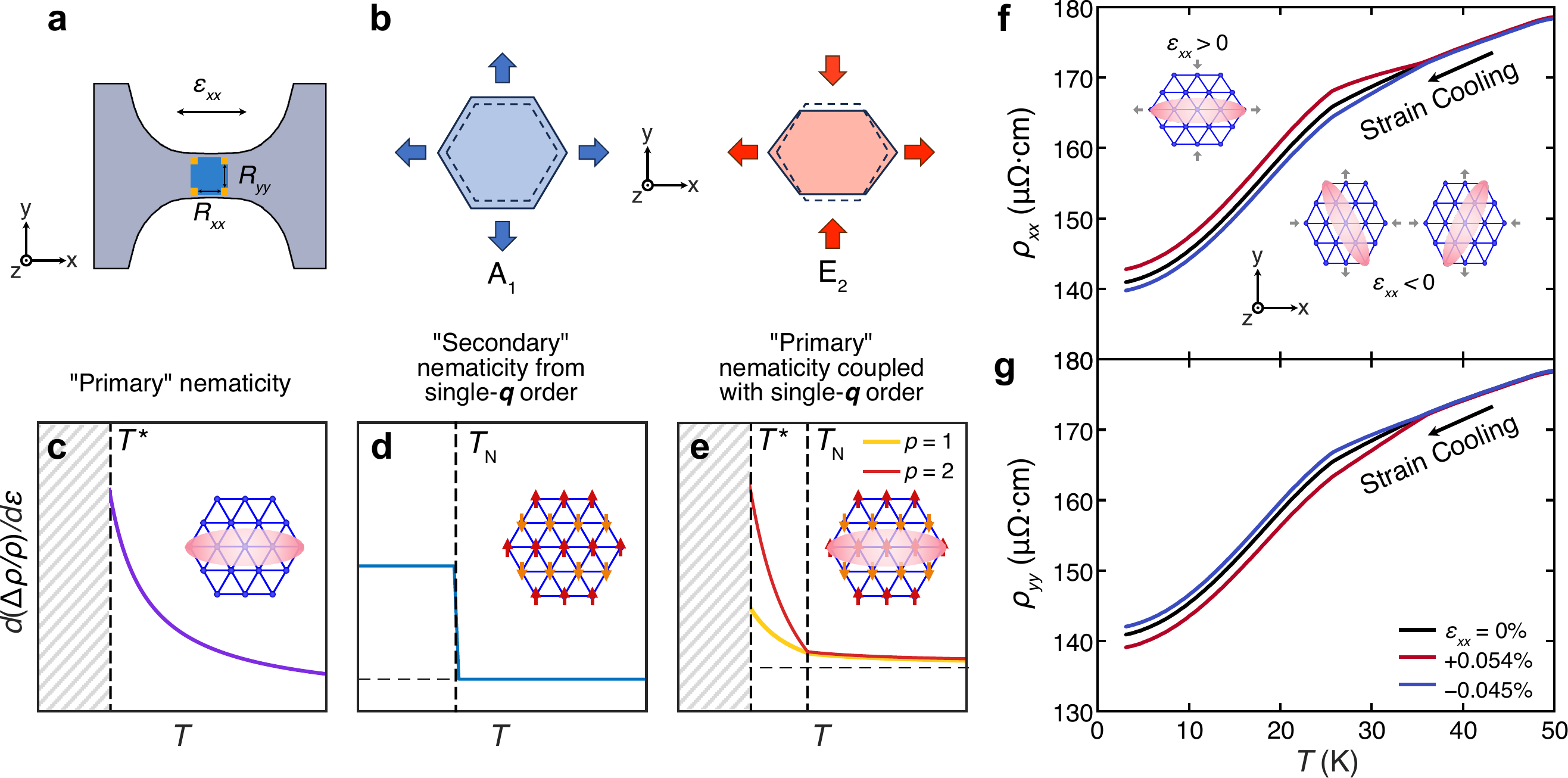}   
    \caption{\textbf{Schematics of probing nematic order in \ce{CoTa3S6} with external strain.}
    (a) Schematic of the modified Montgomery technique for elastoresistivity measurement. A near square sample with four contacts on the corner is glued onto a titanium platform to apply uniaxial strain along the $x$-direction, with resistances $R_{xx}$ and $R_{yy}$ measured at the same time, which are converted to resistivities $\rho_{xx}$ and $\rho_{yy}$ (see Methods). (b) Schematic of strain deformations classified by the $A_{1}$ (left) and $E_{2}$ (right) irreducible representations in the $D_{6}$ point group; throughout the work we adopt a Cartesian coordinate system ($x$, $y$, $z$). (c) Landau-theory-based schematic temperature dependence of nematic susceptibility ($d[\Delta\rho/\rho]/d\epsilon$) for a primary nematicity scenario, where nematicity exists as an independent order parameter. The inset shows the nematic order on the triangular lattice, where the long axis of the red shaded ellipse correspond to the direction along which $\Delta \rho >0$. (d) Schematic temperature dependence of nematic susceptibility ($d[\Delta\rho/\rho]/d\epsilon$) for a secondary nematicity scenario induced by single-$\boldsymbol{q}$ (e.g. AFM) order, in which nematicity arises as a ``by product" associated with magnetic symmetry breaking. $T_N$ corresponds to the single-$\boldsymbol{q}$ (e.g. AFM) transition temperature (see text). The inset shows a typical single-$\boldsymbol{q}$ AFM order on the triangular lattice.
    (e) Corresponding schematic for a primary nematicity scenario coupled to single-$\boldsymbol{q}$ (e.g. AFM) order, where $p=1,2$ demonstrates the results from different nemato-magnetic coupling strength (see Methods). 
    In this model, we adopt the case where AFM transition temperature $T_N$ is above $T^*$. Below $T^*$, hatched area indicates the hysteretic region with multiple nematic domains.
    (f-g) Temperature dependence of the longitudinal resistivity $\rho_{xx}$ and $\rho_{yy}$ showing strain-cooling–induced selection of nematic domains in \ce{CoTa3S6} under tensile (red curve) and compressive strain (blue curve), corresponding to the domain configurations illustrated schematically.}
    \label{fig:figure1}
\end{figure*}

Electronic nematicity refers to a state that spontaneously breaks the discrete lattice rotational symmetries where translational symmetries are preserved. It has emerged as a widespread phenomena in correlated electron systems \cite{fradkin2010nematic}, and frequently found to be intertwined with other ordered phases--such as antiferromagnetism, charge density wave and superconductivity. Historically, nematic phases were first identified in ultra-clean quantum Hall systems \cite{lillyEvidenceAnisotropicState1999,duStronglyAnisotropicTransport1999}, while they are later reported to play a key role in iron-based superconductors, where the superconducting dome emerges in the vicinity of the quantum critical points of stripe magnetism and nematic orders, pointing to a close connection between nematicity and unconventional superconductivity  \cite{Chu2010-cl,Fernandes2014-qg}. More recently, nematic orders have been reported in more extensive families of quantum materials, including nickel analogues of iron-based superconductors \cite{Eckberg2020-so}, twisted multi-layer graphene \cite{Cao2021-sv,Fernandes2020-us}, kagome metals \cite{Li2023-ag, farhangDiscoveryIntermediate2025, Xu2022-hu} and superconductors \cite{bohmerNematicityNematic2022, Borzi2007-li, Hinkov2008-dh, Chu2010-cl}, and correlated low dimensional magnets \cite{little2020three, hwangbo2024strain, tanObservationThreeStateNematicity2024, Yao2025Potts}, motivating a more comprehensive picture of underlying mechanisms leading to the ubiquitous presence of nematic phases across otherwise disparate material platforms.

From an experimental perspective, nematic order is often identified through the emergence of in-plane anisotropy below an ordering temperature in macroscopic probes such as electronic transport \cite{Chu2010-cl,Cao2021-sv} and optical response coefficients \cite{Xu2022-hu}, or alternatively through the observation of rotational-symmetry breaking in momentum-space spectroscopies \cite{Li2023-ag,farhangDiscoveryIntermediate2025}. However, many rotational symmetry-breaking patterns are, in principle, also compatible with ordered states that involve additional translational symmetry breaking. A common example is the development of a finite modulation wave vector $\mathbf{q}$ that selects a preferred spatial direction, as realized in stripe magnetic orders \cite{little2020three} and unidirectional charge- or spin-density-wave phases \cite{chenUnidirectionalSpin2018,guoSpectralEvidence2023}. Indeed, nematic order can be intertwined with such states as a vestigial phase \cite{Fernandes2019-kt}, underscoring the need for experimental tools capable of distinguishing genuine nematic order—where rotational symmetry breaking is primary—from spatially modulated phases in which it appears as a secondary consequence.
In this context, symmetry-breaking strain, together with associated transport and thermodynamic probes \cite{Chu2012-vj,bohmer2022nematicity,doi:10.1073/pnas.2105911118}, has been increasingly established as an effective conjugate field to the nematic order parameter and its fluctuations, owing to the fact that strain breaks identical symmetries as a primary nematic order and thereby couples bilinearly to it. In particular, elastoresistivity has emerged as a powerful technique for detecting nematic fluctuations and for quantitatively characterizing the strain–nematic coupling \cite{Chu2012-vj,kuoMeasurementElastoresistivity2013,kuo2016ubiquitous,rosenbergDivergenceQuadrupolestrain2019,Eckberg2020-so,Frachet2022-bt}.

In this work, we will apply elastoresistivity to explore the nature of a recently reported system to host the nematic order, namely the layered antiferromagnet \ce{CoTa3S6}. \ce{CoTa3S6} belongs to a family of magnetic intercalated transition metal dichalcogenides \cite{parkin19803A,miyadai1983magnetic,morosan2007sharp,wu2022highly,xie2022structure}, and renewed interest in this class of materials has been driven by the discovery of a large anomalous Hall effect within the antiferromagnetic state of \ce{CoTa3S6} and its isostructural cousin \ce{CoNb3S6} \cite{Ghimire2018-bn,park2022field,PhysRevB.103.184408,PhysRevResearch.2.023051,tanaka2022large}. Two successive magnetic transitions are reported in \ce{CoTa3S6} at $T_{N1}\approx37~\mathrm{K}$ and $T_{N2}\approx26 ~\mathrm{K}$, with anomalous Hall effect observed below $T_{N2}$ \cite{parkin19803A,park2022field}. Neutron scattering experiments suggest that the magnetic structure on the \ce{Co} triangular lattice is a single-$\boldsymbol{q}$ collinear antiferromagnetic order right below $T_{N1}$, which then turns into a non-coplanar triple-$\boldsymbol{q}$ structure below $T_{N2}$ \cite{park2023tetrahedral,takagi2023spontaneous,Park2024}. The latter state is expected to host finite scalar spin chirality to account for the topological origin of the observed anomalous Hall effect \cite{park2023tetrahedral,takagi2023spontaneous,yanagi2023generation,park2024dft+,heinonen2022magnetic}. More recently, in addition to time-reversal symmetry breaking, rotational symmetry breaking is reported in the system, where in-plane resistivity anisotropy and optical birefringence emerge below $T_{N1}$ at $T^*\simeq 34\mathrm{~K}$ and persist down to low temperature \cite{fengNonvolatileNematic2025}. The appearance of electronic anisotropy at a temperature between $T_{N1}$ and $T_{N2}$—and its survival into the triple-$\boldsymbol{q}$ ground state—is unexpected and suggests an origin distinct from the magnetic phase transitions. Clarifying the origin of the anisotropy, and how this anisotropy is related to and distinct from underlying magnetic orders remains an open question that motivates the present study.

 \begin{figure*}[t]
    \centering
   \includegraphics[width=0.98\textwidth]{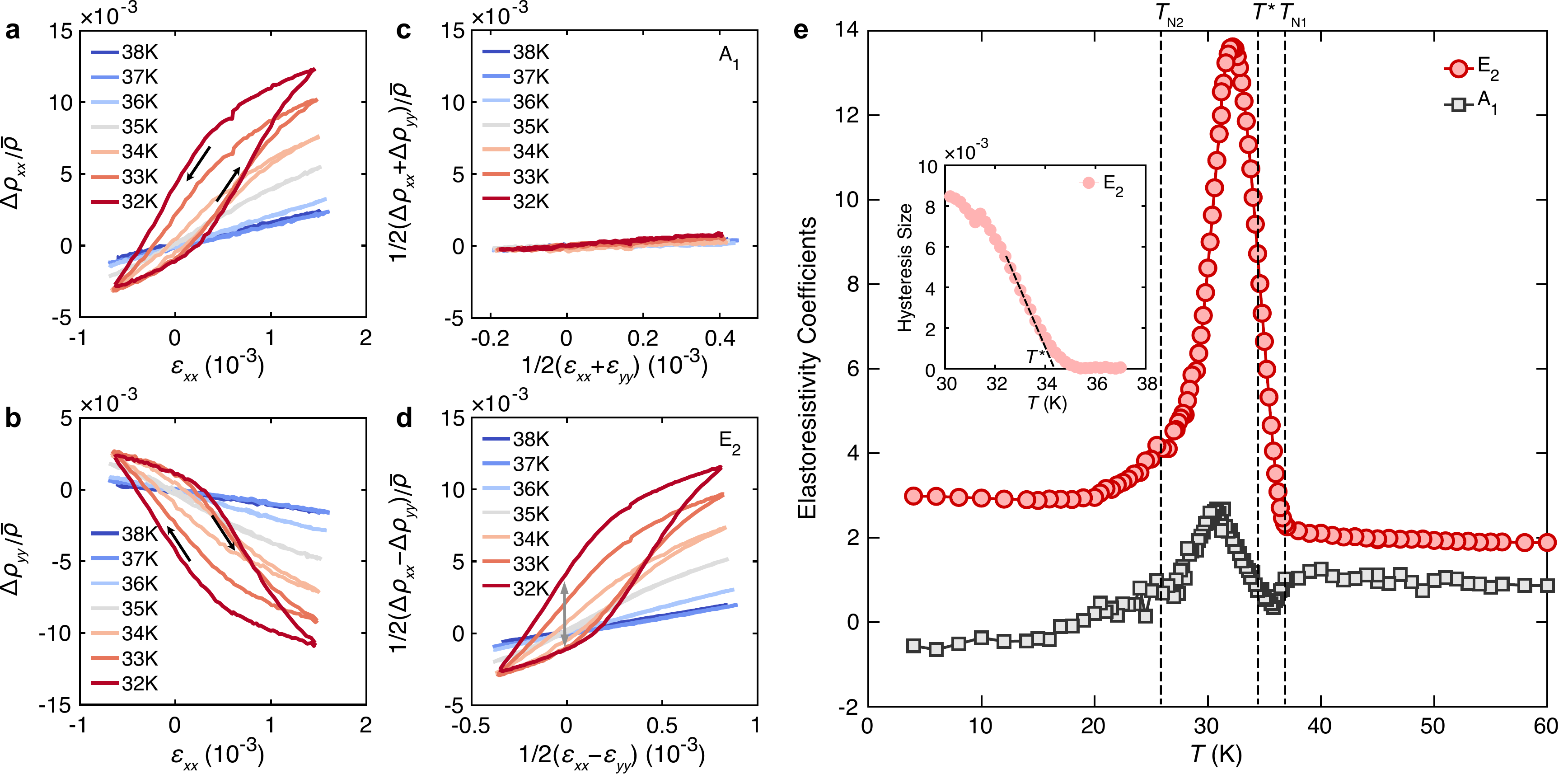}
    \caption{\textbf{Elastoresistivity and nematic susceptibility.}
    (a-b) Relative changes of resistivity $\rho_{ii}$ ($i=x,y$) as a function of strain (defined as $\Delta \rho_{ii}=\rho_{ii}-\bar{\rho}$) at representative temperatures measured by the modified Montgomery method. $\bar{\rho} = (\rho_{xx,0}+\rho_{yy,0})/2$ is the average in-plane resistivity, where $\rho_{xx,0}$ and $\rho_{yy,0}$ are the resistivity value at $\varepsilon_{xx}=0$ for the increasing strain scan. The black arrows denote the strain scan direction for the hysteresis loop. (c-d) Representative data of resistivity decomposed to the isotropic $A_{1}$ and anisotropic $E_{2}$ channels at different temperatures. The gray arrow in (d) defines the size of hysteresis at $\varepsilon = 0$. (e) Temperature dependence of elastoresistivity coefficients (see Methods). Transition temperatures $T_{\mathrm{N1}}$, $T^*$ , $T_{\mathrm{N2}}$ are indicated with dashed lines. Inset shows the temperature dependence of hysteresis size of $E_{2}$ channel defined in (d) in the resistivity-strain scan. $T^*$ is defined as the onset temperature of hysteresis: intercept of the dashed line with the temperature axis. }
    \label{fig:figure2}
\end{figure*}

Here we carry out a comprehensive elastoresistivity study on \ce{CoTa3S6} to address the nature of the observed anisotropic nematic order. The experimental configuration is shown in Fig. \ref{fig:figure1}a, where the longitudinal strain $\varepsilon_{xx}$ along the titanium bridge is controlled experimentally in an in-situ manner with resistivities $\rho_{xx}$ and $\rho_{yy}$ measured simultaneously, allowing one to probe multiple components of the elastorestivity tensor $m_{ij}$ (see Methods). 
We note that the strained sample shown in Fig. \ref{fig:figure1}a will be subject to both symmetric (belonging to the $A_1$ irreducible representation of the $D_6$ point group) and antisymmetric (belonging to the $E_2$ irreducible representation of $D_6$) strain as illustrated in Fig. \ref{fig:figure1}b, allowing one to extract the proper irreducible components of the elastoresistivity tensor $m_{A_1}$ and $m_{E_2}$ (see Methods) where the latter $m_{E_{2}}
= d\left[(\Delta \rho / \rho)_{xx}-(\Delta \rho / \rho)_{yy}\right]/
       d\left[\epsilon_{xx}-\epsilon_{yy}\right]$ is expected to be directly capturing the susceptibility and fluctuations of a nematic order that breaks in-plane rotation symmetries. Here $\Delta\rho$ is defined as the deviation from the corresponding isotropic resistivity $\bar{\rho}$ with $\Delta\rho_{ii}=\rho_{ii}-\bar{\rho}$.

In Fig. \ref{fig:figure1}c-e, we illustrate three different scenarios for the antisymmetric elastoresistance coefficient $d[\Delta\rho/\rho]/d\epsilon$ within a Landau free energy framework: a pure nematic order with order parameter $\eta_n$ (Fig.~\ref{fig:figure1}c, with $\Delta\rho/\rho\propto\eta_n$), `secondary' nematicity driven by the emergence of a finite-$q$ vector $\eta_q$ (Fig.~\ref{fig:figure1}d, with $\Delta\rho/\rho\propto\eta^2_q$), and a third scenario where $\eta_n$ and $\eta_q$ are distinct primary order parameters which coexist (Fig.~\ref{fig:figure1}e, with $\Delta\rho/\rho\propto\eta_n$). The crucial difference among these scenarios is that the leading symmetry-allowed order parameter-strain coupling takes the form $\eta_n\epsilon$ for a genuine nematic order and the form of $\eta_q^2\epsilon$ for the `secondary' rotation symmetry-breaking.
In the pure nematic case  (Fig. 1c), above the ordering temperature, a Curie-Weiss divergence is expected in $d[\Delta\rho/\rho]/d\epsilon$ analogous to the magnetic susceptibility near a ferromagnetic order \cite{Chu2012-vj, bohmer2022nematicity, riggsEvidenceNematic2015}; below the ordering temperature $d[\Delta\rho/\rho]/d\epsilon$ becomes poorly defined as strain hysteresis onsets. 
 In the second scenario (Fig.~\ref{fig:figure1}d), no linear strain-driven anisotropy is expected above the transition (corresponding to $d[\Delta\rho/\rho] /d\epsilon=0$), and $d[\Delta\rho/\rho]/d\epsilon$ shows a step-wise increase at the transition \cite{luthiPhysicalAcoustics2007, sunIntertwinedChargeDensity2025}. In the third scenario, aside from the onset of separate orders, terms in the form of $\eta_q^2\eta_n$ and $p\eta_q^2\eta_n\epsilon$ characterizing the interplay between $\eta_n$ and $\eta_q$ are additionally allowed. A sizable $p\eta_q^2\eta_n\epsilon$ coupling can lead to a marked increase in the elastoresistivity coefficient below $T_{N}$ on top of a Curie-Weiss behavior approaching $T^*$ (Fig.~\ref{fig:figure1}e, see Methods and Supplementary Materials).

\noindent\textbf{Strain response and elastoresistivity coefficients of \ce{CoTa3S6}} In Fig.~\ref{fig:figure1}f(g), we show $\rho_{xx}$($\rho_{yy}$) along $a$($b^*$)-axis for a \ce{CoTa3S6} sample on the strain platform cooled under zero (black), moderate tensile (red) and compressive (blue) $\epsilon_{xx}$. While both resistivities exhibit negligible strain dependence at high temperatures, a pronounced and opposite strain response in $\rho_{xx}$ and $\rho_{yy}$ emerges below $\sim37~\mathrm{K}$. This points to the onset of electronic anisotropy upon entering the magnetic/nematic regime and is consistent with strain-controlled nematic domains at low temperatures, as schematically illustrated in the inset of Fig.~\ref{fig:figure1}f \cite{fengNonvolatileNematic2025}. 

\begin{figure*}[th]
    \centering    \includegraphics[width=0.95\linewidth]{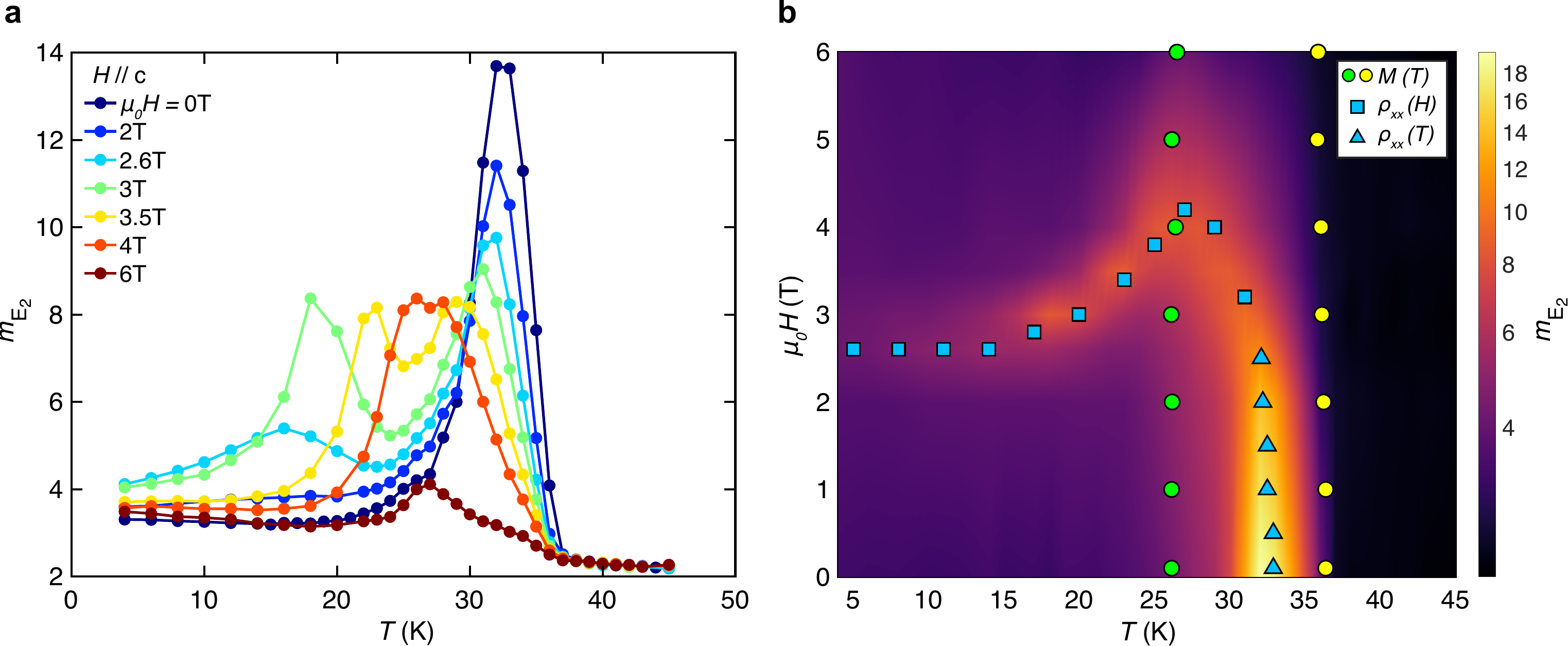}
    \caption{\textbf{Nematic fluctuations controlled by out-of-plane magnetic fields.}
    (a) Temperature dependence of the elastoresistivity coefficient in the $E_{2}$ channel ($m_{E_2}$), measured under different out-of-plane magnetic fields. (b) Colormap of $m_{E_2}$ in log scale, overlaying with magnetic transition temperatures, illustrating the $H$--$T$ phase diagram. The phase boundaries $T^*$ (blue triangles), $H_c$ (blue squares), $T_{N1}$ (yellow dots) and $T_{N2}$ (green dots) are indicated by symbols extracted from anomalies in $M(T)$ (circles), $\rho_{xx}(H)$ (squares), and $\rho_{xx}(T)$ (triangles), compiled from previous work~\cite{fengNonvolatileNematic2025}.
}
    \label{fig:figure3}
\end{figure*}

In Fig.~\ref{fig:figure2}a-b we show normalized $\Delta\rho_{xx}$ and $\Delta\rho_{yy}$ as a function of $\epsilon_{xx}$, which are largely linear above 37 K, while a finite hysteresis gradually develops upon cooling below $\simeq 34\mathrm{~K}$. We attribute the observed hysteresis to the presence of nematic domains at low temperature. After decomposing the elastoresistivity into isotropic $A_{1}$ and anisotropic $E_{2}$ channels (Fig.~\ref{fig:figure2}c-d), we find that $\Delta\rho/\bar{\rho}-\epsilon$ is linear at all temperatures in the $A_{1}$ channel (Fig. 2c); responses in the $E_{2}$ channel (Fig. 2d), in contrast, is highly nonlinear and hysteretic below 35 K. The inset of Fig.~\ref{fig:figure2}e summarizes the size of $E_{2}$ hysteresis (Fig. 2d) at zero strain in $\Delta\rho/\bar{\rho}$, which allows one to define the onset temperature $T^*$ for the nematic order $T^*\simeq 34$ K,  consistent with that determined from magnetic field-trained resistivity \cite{fengNonvolatileNematic2025}. 

The slopes, namely the corresponding elastoresistivity coefficients, are summarized in Fig.~\ref{fig:figure2}e where  $m_{E_2}$ are shown as red and $m_{A_1}$ as black symbols, respectively. $m_{A_{1}}$ exhibits weak temperature dependence over the entire measured temperature range; meanwhile, $m_{E_{2}}$ is at least one order of magnitude larger and exhibits a pronounced peak around 32 K: this contrast highlights that the dominant strain response of the system lies in the anisotropic $E_2$ channel, reflecting an underlying rotational symmetry breaking. 
We note that the temperature evolution of $m_{E_{2}}$ is rather intriguing: above $T_{N1}$, $m_{E_{2}}$ slowly increases with decreasing $T$; an upturn occurs at $T_{N1}$ as $m_{E_{2}}$ approaches $T^*$ divergently. At about 32 K below $T^*$, $m_{E_{2}}$ reaches its maximum and subsequently decreases, which we attribute to a progressive freezing and pinning of nematic domains as thermal fluctuations are suppressed. We note that the measured $m_{E_{2}}$ closely resembles the free energy simulation shown above in Fig.~\ref{fig:figure1}e, in contrast to the step-like enhancement at the transition temperature (Fig.~\ref{fig:figure1}d) as expected and observed in systems where rotational symmetry is simply broken as a by-product of a spatially varying order. Our observations therefore rule out a scenario in which the resistivity anisotropy in \ce{CoTa3S6} is merely a consequence of the collinear, single-$q$ antiferromagnetic order; instead, they are consistent with a primary nematic order that couples bilinearly to external strain which further intertwines with the collinear antiferromagnetic state.\\

\begin{figure*}
    \centering
    \includegraphics[width=0.9\linewidth]{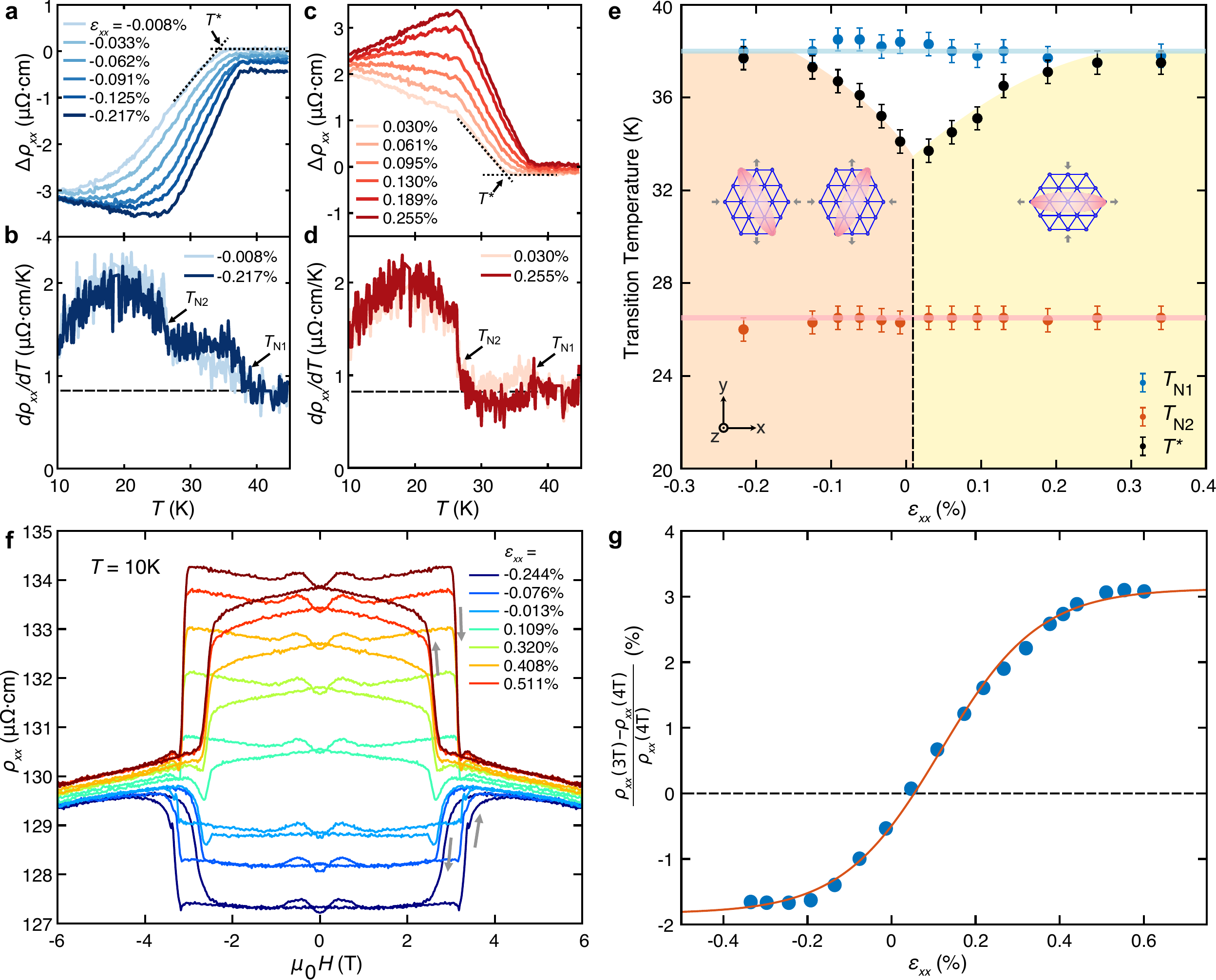}
    \caption{\textbf{Strain dependence of transition temperatures and features of three-state nematicity.}
    (a,c) Resistivity anisotropy $\Delta \rho_{xx}$ obtained by subtracting a zero strain background from strained states (see Methods), from which $T^*$ under different bias strains is determined. (b,d) Representative $d\rho_{xx}/dT$ curves under bias strains where $T_{N1}$ and $T_{N2}$ are determined from kinks. (e) Summarized strain dependence of the transition temperatures $T_{N1}$, $T_{N2}$ and $T^*$. $T_{N1}$ ($T_{N2}$) under different bias strains fall on the line of $T=38\mathrm{~K}$ (blue) ($T=26.5\mathrm{~K}$ (red)). Two shaded regions divided by the dashed line indicate the nematic phases with illustrated domains favored by compressive (left) and tensile (right) strains respectively. (f) Out-of-plane magnetic field ($H\parallel z$) dependence of $\rho_{xx}$ measured at 10 K under various magnitudes of strain along the $x$ axis ($\epsilon_{xx}$). (g) Strain dependence of nematic order parameter at each bias strain projected on the $x$-axis extracted from $(\rho_{xx}(3\mathrm{T})-\rho_{xx}(4\mathrm{T}))/\rho_{xx}(4\mathrm{T})$ (see text). The red line is a fit to domain population formula modified for the three-state nematicity (see SI).}
    \label{fig:figure4}
\end{figure*}

\noindent\textbf{Nematic fluctuations modulated by out-of-plane magnetic field} In order to gain additional insights on the interplay between nematicity and the underlying magnetic orders in \ce{CoTa3S6}, below we trace the effect of $c$-axis magnetic field to the measured nematic susceptibility $m_{E_2}$, recalling that previous works reported that the nematic order can be fully suppressed by a moderate out-of-plane magnetic field \cite{fengNonvolatileNematic2025}. Fig.~\ref{fig:figure3}a shows $m_{E_{2}}-T$ under selected out-of-plane magnetic field $H$: with increasing $H$, the peak of $m_{E_{2}}$ is gradually suppressed and shifts towards lower temperatures. Above $\mu_0H = 2.6~\mathrm{T}$, an additional peak develops on the low-temperature side $\sim16~\mathrm{K}$. The second peak gradually shifts toward higher temperatures with increasing $H$, and merges with the higher temperature peak at $\mu_0H \simeq 4~\mathrm{T}$. With further increasing $H$, the overall magnitude of $m_{E_{2}}$ is suppressed and becomes nearly temperature independent.

Figure~\ref{fig:figure3}b presents a color map of $m_{E_2}$ in the $T-H$ plane. The brightest ridge in the color map marks the locations where the nematic fluctuations are strongest. We have overlaid the phase boundaries $T_{N1}, T_{N2}$ and $T^*$ (compiled from previous work~\cite{fengNonvolatileNematic2025}) with the color map, where one can see that the nematic fluctuations are strongest at zero field right below $T^*$, and at finite magnetic fields the enhanced nematic fluctuations directly follows the field-evolution of the phase boundary defined by $T^*$. That the phase boundary intersects the field axis at $\mu_0H_c \approx 2.6~\mathrm{T}$ in the low temperature limit \cite{fengNonvolatileNematic2025} is also captured in constant temperature cuts of $m_{E_2}$ (see SI), and the double peak structure in the $T$-scans at intermediate magnetic fields is consistent with a non-monotonic $T$-evolution of $H_c$. These observations therefore highlight that the nematic fluctuations in \ce{CoTa3S6} are sensitively captured by elastoresistance and are highly tunable by out-of-plane magnetic field. \\

\noindent\textbf{Strain evolutions of the nematic order} In the following we examine the strain evolution of the nematic and antiferromagnetic critical temperatures $T^*$, $T_{N1}$ and $T_{N2}$. Here $T^*$ is defined via $\Delta\rho_{xx}$ obtained by subtracting a zero strain background $\rho_{xx}-T$ from strained states (see Methods and Fig.~\ref{fig:figure4}a and c) while $T_{N1}$ and $T_{N2}$ are determined from kinks in $d\rho_{xx}/dT$ (Fig. 4b and d). We summarize the strain-dependence of  $T^*$, $T_{N1}$ and $T_{N2}$ in Fig. \ref{fig:figure4}e as black, blue and red symbols, respectively. While $T_{N1}$ and $T_{N2}$ remain nearly unchanged over the entire strain range in the study (from $-0.3\%$ to $+0.4\%$), the nematic onset temperature $T^*$ exhibits a pronounced `V-shape' strain dependence, which could be attributed to two sets of different strain-selective domains as illustrated in the inset of Fig. \ref{fig:figure4}e. Similar `V-shape' strain dependence for different domains have been reported in multiple systems with rotational symmetry breaking \cite{Straquadine2022, Ye2023,hwangbo2024strain}. Specifically, $T^*$ increases from  $\sim34~\mathrm{K}$ at near-zero bias strain to about $38\mathrm{~K}$ for strains exceeding $0.2\%$ on either the tensile or compressive side, where $T^*$ coincides and appears to merge with $T_{N1}$.

 As \ce{CoTa3S6} hosts a hexagonal crystalline lattice, the nematic order in the system in principle contains three inequivalent domains and can be mapped onto the three-state Potts model, in contrast to the two-state Ising nematicity in tetragonal systems \cite{Chu2010-cl,rosenbergDivergenceQuadrupolestrain2019}. Here we present signatures of the three-state nature of the  nematicity in \ce{CoTa3S6}.  In Fig.~\ref{fig:figure4}f we show $\rho_{xx}$ as a function of out-of-plane magnetic field at $T = 10~\mathrm{K}$ at selected bias strains. The sign switching of magnetoresistance from tensile (negative) to compressive (positive) strain is consistent with Ref.  \cite{fengNonvolatileNematic2025}. Given that at around $\mu_0H = 3~\mathrm{T}$ the system evolves from the nematic phase to a $C_3$ rotational symmetric phase, we quantify the nematic order parameter at each bias strain projected on the $x$-axis as the jump in $\rho_{xx}$ at $H_c$: $\Delta\rho_{xx}/\rho_0(H_c)$. This relies on the assumption that the high-field $\rho_{xx}$ provides a good approximation to a corresponding multi-domain, isotropic state at the same bias strain. Specifically, we define $\Delta\rho_{xx}/\rho_0(H_c)\equiv \frac{\rho_{xx}(3\mathrm{T})-\rho_{xx}(4\mathrm{T})}{\rho_{xx}(4\mathrm{T})}$. 

In Fig.~\ref{fig:figure4}g we showcase $\Delta\rho_{xx}/\rho_0(H_c)$ as a function of bias strain. For the three state Potts model, the free energy could be written as 
$
F_{\mathrm{n}}=b_0\left(T-T_n\right) n^2+c n^3 \cos 6 \theta+w n^4-\lambda n \epsilon \cos \left(2 \theta-2 \theta_S\right)
$,
where $\boldsymbol{n} = n(\cos\theta, \sin\theta)$ is the nemtaic order parameter, and $\lambda$ is its coupling strength to external strain $\boldsymbol{\epsilon} = \epsilon(\cos\theta_S, \sin\theta_S)$. The inequivalence for tensile and compressive strain due to the $c n^3 \cos 6 \theta$ term is a key characteristic feature for three-states nematicity, while an Ising nematic system should behave in a symmetric manner on tensile and compressive strains. By minimizing the free energy, we find that the saturation value of $\Delta\rho_{xx}/\rho_0 \propto n\cos2\theta$ under tensile strain (favoring $\theta=0$ domains) should be twice that under compressive strain (favoring $\theta=\pi/3,~2\pi/3$ domains, see Supplementary Information), consistent with our observations in Fig.~\ref{fig:figure4}g. \\

\noindent\textbf{Discussion} 
In this work, using anisotropic elastoresistance, we examine the response of intertwined nematic and magnetic orders in \ce{CoTa3S6} to external (out-of-plane) magnetic fields and uniaxial strains. Our observation of highly field-sensitive nematic fluctuations in elastoresistivity, viewed together with the largely field-independent antiferromagnetic transitions $T_{N1}$ and $T_{N2}$, suggests that the antiferromagnetic orders (in particular the collinear antiferromagnetic order) are distinct from the nematic order which drives the electronic anisotropy in the system. This scenario is further corroborated by the distinct strain responses of $T^*$ and the magnetic transition temperatures.

The elastoresistivity of \ce{CoTa3S6} exhibits a distinct temperature evolution compared to materials in which rotational symmetry breaking is believed to arise as a secondary effect of a single-$\boldsymbol{q}$ state selecting a preferred spatial direction. A representative hexagonal system is the kagome metal \ce{CsV3Sb5}, where recent elastoresistance measurements revealed jump-like anomalies in both $E_{2g}$ and $A_{1g}$ channels, with the response dominated by the $A_{1g}$ signal \cite{sunIntertwinedChargeDensity2025,liu2024absence}. In that case, the reduced rotational symmetry is interpreted as arising from the stacking pattern of the triple-$\boldsymbol{q}$ CDW rather than from a primary electronic nematic instability. In contrast, the elastoresistivity response in \ce{CoTa3S6} evolves continuously below $T_{N1}$ and across $T^*$, and is best described within a free-energy framework involving a nematic degree of freedom that couples bilinearly to symmetry-breaking strain, with the detailed temperature dependence further influenced by the intertwined magnetic order parameters. More generally, our results illustrate that elastoresistivity, when analyzed within a symmetry-based free-energy framework and combined with controlled tuning parameters such as magnetic field and strain, provides a powerful approach for identifying the origin of rotational symmetry breaking and disentangling intertwined ordered states.

\subsection{Acknowledgement}

The experimental works carried out at Caltech are supported by Gordon and Betty Moore Foundation through Moore Materials Synthesis Fellowship to L.Y. (GBMF12765) and Institute of Quantum Information and Matter (IQIM), an NSF Physics Frontier Center (PHY-2317110).  M.Y. is supported by a start-up grant from the University of Utah. M.Y. and L.Y. acknowledge support from the Gordon and Betty Moore Foundation’s EPiQS Initiative (Grant GBMF11918), which enabled valuable discussions.
 Z.F. acknowledges support from IQIM Postdoctoral Fellowship at Caltech. Part of the work at Caltech is carried out at the X-Ray Crystallography Facility supported by the Beckman Institute and at the shared Physical Properties Measurement System facilities supported by NSF DMR-2117094.

\section{Methods}
\subsection{Crystal Growth and Characterization}
Single crystals of \ce{CoTa3S6} were grown by a two-step process reported in \cite{fengNonvolatileNematic2025}.
The typical dimension of the obtained crystals is $1-2~\mathrm{mm}$. The phase purity of the crystals was checked by powder X-ray diffraction, and the orientation of the single crystalline samples was determined by a Laue diffractometer.
Energy dispersive X-ray spectroscopy was used to characterize the resulting stoichiometry of the single crystals.
It was measured on the ZEISS 1550VP field emission SEM with Oxford X-Max SDD X-ray Energy Dispersive Spectrometer (EDS) system.

\subsection{Modified Montgomery Method}
The modified Montgomery method is an effective technique for measuring the anisotropic resistivity in orthorgonal directions within a single sample simultaneously \cite{montgomery1971method,dos2011procedure,liu2024absence}. 
In this study, \ce{CoTa3S6} is polished into a thin, nearly square shape with its edges aligned along the  $x$- and  $y$-axes, respectively.
Four electrodes were attached at the corners of the sample with relative small contact area to ensure precise measurements.
The resistance $R_{xx}$  ($R_{yy}$) was determined by applying current $I_x$ ($I_y$) along the $x$-($y$-) axis on one side and measuring the voltage drop $V_x$ ($V_y$) on the opposite side; and $R_{xx}=V_x/I_x$ ($R_{yy}=V_y/I_y$).
The resistivities $\rho_{xx}$  and  $\rho_{yy}$ were calculated based on the ratio of  $R_{xx}$  and  $R_{yy}$  (see the Supplementary Information for details). 

\subsection{Landau Free Energy Modeling}
The schematic temperature dependence of nematic susceptibility $d[\Delta\rho/\rho]/d\epsilon$ is modeled within a framework based on Landau theory. For simplicity, we consider a reduced phenomenological model in which the nematic order parameter is projected onto a single component along the selected symmetry-breaking direction ($\theta = 0$). For a primary nematicity scenario, a pure nematic order is described by an order parameter $\eta_n$, and the free energy of the system is 
\begin{equation}
F_n=\frac{1}{2} a\left(T-T_0\right) \eta_n^2+\frac{1}{4} b \eta_n^4-g \epsilon \eta_n,
\end{equation}
Solving $dF_n/d\eta_n = 0$ results in the Curie-Weiss behavior $d\eta_n/d\epsilon \propto g/[a(T-T_0)]$, as plotted in Fig. \ref{fig:figure1}c. Higher-order symmetry-allowed terms (such as cubic terms relevant for three-state nematicity) do not modify the leading Curie–Weiss behavior of the nematic susceptibility and are therefore omitted here for clarity.

For a secondary nematicity scenario from a single-$\boldsymbol{q}$ order, the order parameter $\eta_q$ quadratically couples to strain as $\eta_q^2\epsilon$, and
\begin{equation}
F_q=\frac{1}{2} a\left(T-T_0\right) \eta_q^2+\frac{1}{4} b \eta_q^4-g \epsilon \eta^2_q,
\end{equation}
Solving $dF_q/d\eta_q = 0$ results in
\begin{equation}
\frac{d\left(\eta_q^2\right)}{d \epsilon}= \begin{cases}0, & T>T_0 \\ 2 h / b, & T<T_0\end{cases},
\end{equation}
as plotted in Fig. \ref{fig:figure1}d. 

For the case of a primary nematic order coupled to a single-$\boldsymbol{q}$ order (including the cubic term), the free energy writes:
\begin{equation}
F = b_0(T-T_n) \eta_n^2 + c \eta_n^3 + w \eta_n^4 -(\lambda+p\eta_q^2) \eta_n\epsilon - g \eta_q^2\eta_n,
\end{equation}
where $\lambda_{eff} = \lambda+p\eta_q^2$ is the renormalized nemato-elastic coupling dependent on the single-$\boldsymbol{q}$ order (e.g. magnetization), and $p$ reflects the nemato-magnetic coupling strength.
The free energy $F$ is minimized with respect to $\eta_n$ numerically by choosing the values of parameters as follows: $T_n=30$, $b_0=2$, $c=-0.5$, $w=5$, $\lambda=10$, $p=1,2$, $g = 0.1$, bias strain $\epsilon_{bias} = 0.005$; for the single-$\boldsymbol{q}$ order, we use the simplest mean-field result $\eta^2_q=k\left|T_{N1}-T\right|$, where $k=3$ and $T_{N1}=37$. The resulting $d\eta_n/d\epsilon$ as a function of $T$ is plotted in Fig. \ref{fig:figure1}e.

\subsection{Transport Measurement under Strain}
Electrical transport properties were measured using both standard four-wire methods and modified Montgomery method in the commercial cryostat TeslatronPT (Oxford Instruments) with a superconducting magnet. Strain was applied to single crystals of \ce{CoTa3S6} using a commercial piezoelectric uniaxial strain cell (CS-100, Razorbill Instruments). For data shown in Fig.2, a near square-shaped sample ($0.80\times0.86\times 0.02$ mm$^3$) is affixed onto a titanium platform in bowtie shape (Fig.~\ref{fig:figure1}a, thickness $0.3\mathrm{~mm}$), which is later attached to CS-100 cell to generate strain (see dedicated section below). For data shown in Fig. 4, a polished bar-shape sample with typical dimension of $1.5\times0.68\times 0.1$ mm$^3$ was mounted to the strain cell with Stycast 2850FT epoxy. The center part of the sample between the mounting plates is approximately 0.7 mm -- we assume this to be the length $L$ of the strained region of the sample and it is used to estimate the applied strain through $\epsilon_{xx}=\Delta L/L$ where $\Delta L$ is obtained from the capacitive displacement sensor in the strain cell. The piezoelectric stacks were controlled by a combination of the DC output of Keithley 6221 current source and TEGAM 2350 high voltage amplifier.  Longitudinal and Hall resistivities were measured using external lock-in amplifiers (Stanford Research Systems, SR860) with an AC current of typical amplitude 2 mA from a voltage controlled current source (Stanford Research Systems, CS580).

\subsection{Elastoresistivity Tensor in D6 Point Group}
Elastoresistivity is a fourth-rank tensor that can be expressed as $m_{i j}=\frac{\partial(\Delta \rho / \rho)_i}{\partial \epsilon_j}$, where $i$ and $j$ take values from 1 to 6 corresponding to $xx$, $yy$, $zz$, $yz$, $zx$ and $xy$, following the Voigt notation. These elastoresistivity tensors can be further grouped and decomposed into different symmetry channels based on the irreducible representation from the crystal lattice point group. In the $D_{6}$ point group as in the case of \ce{CoTa3S6}, the nematic susceptibility corresponds to the elastoresistivity coefficient $m_{E_{2}}= m_{11}-m_{12}$, and the symmetric mode with $A_{1}$ symmetry corresponds to $m_{A_{1}}= m_{11}+m_{12}-m_{13}[2\nu_{ac}/(1-\nu_{ab})]$, where $\nu_{ac,ab}$ are the out-of-plane and in-plane Poisson ratios respectively. A near square-shaped sample is affixed onto a titanium platform in bowtie shape, which is later attached to a piezoelectric strain cell (CS-100, Razorbill) to apply uniaxial strain onto the sample and generate a combination of anisotropic $\frac{1}{2}(\epsilon_{11}-\epsilon_{12})$ and isotropic $\frac{1}{2}(\epsilon_{11}+\epsilon_{12})$ strain components. We use the modified Montgomery methods to measure $\rho_{xx}$ and $\rho_{yy}$ simultaneously and extract the strain-induced change in resistivity $d[(\Delta \rho / \rho)_{xx}]/d\epsilon_{xx}$ and $d[(\Delta \rho / \rho)_{yy}]/d\epsilon_{xx}$ from the linear fitting of $(\Delta \rho_{xx} / \bar{\rho})-\epsilon_{xx}$ (Fig. \ref{fig:figure2}a) and $(\Delta \rho_{yy} / \bar{\rho})-\epsilon_{xx}$ (Fig. \ref{fig:figure2}b) scans. In the hysteretic region, the resistance values from the two scan directions at the same strain are averaged before performing a linear fit over the strain range from $-0.04\%$ to $+0.12\%$. And the corresponding elastoresistivity tensors are calculated using the following formula:

\begin{equation}
\begin{aligned}
m_{E_{2}}
&= \frac{d\left[(\Delta \rho / \rho)_{xx}-(\Delta \rho / \rho)_{yy}\right]}
       {d\left[\epsilon_{xx}-\epsilon_{yy}\right]} \\
&= \frac{1}{1+\nu_{\mathrm{p}}}
  \frac{d\left[(\Delta \rho / \rho)_{xx}-(\Delta \rho / \rho)_{yy}\right]}
       {d \epsilon_{xx}} ,
\end{aligned}
\end{equation}
\begin{equation}
\begin{aligned}
m_{A_{1}}
&= \frac{d\left[(\Delta \rho / \rho)_{xx}+(\Delta \rho / \rho)_{yy}\right]}
       {d\left[\epsilon_{xx}+\epsilon_{yy}\right]} \\
&= \frac{1}{1-\nu_{\mathrm{p}}}
  \frac{d\left[(\Delta \rho / \rho)_{xx}+(\Delta \rho / \rho)_{yy}\right]}
       {d \epsilon_{xx}} .
\end{aligned}
\end{equation}
Here $\nu_p\approx0.33$ is approximated as the Poisson ratio of the titanium platform since the sample is much thinner.

\subsection{Resistivity Anisotropy under Applied Bias Strains}
The resistivity anisotropy (proxy for nematic order parameter) as a function of temperature under different bias strains in Fig. \ref{fig:figure4}a and c are taken in the following sequence. The system is cooled across $T^*$ with no external strain and no magnetic field applied. Then a warming scan of $\rho_{xx}(T)$ from $10\mathrm{~K}$ to $45\mathrm{~K}$ is recorded and use this as the zero-strain background. To prepare a state with finite macroscopic nematic order parameter, the sample is first cooled from $45\mathrm{~K}$ to $10\mathrm{~K}$ with no strain applied. Then the out-of-plane magnetic field is set to $6\mathrm{~T}$ at $10\mathrm{~K}$, much above the critical field that restores the rotational symmetry. Then an external tensile (compressive) strain $\epsilon_{xx}\approx+0.2\%$ ($\epsilon_{xx}\approx-0.2\%$) is applied before magnetic field sweeping back to $0\mathrm{~T}$. This process is defined as an in-situ strain training assisted by the field-induced transition at $H_c$ (see SI), which is better than selecting the domain by cooling the system with an external strain (more precisely, a fixed voltage on the piezo stack), because the latter would substantially shift the zero-strain condition due to the temperature- and voltage-history dependence of the Razorbill piezo stack. After this in-situ strain training process, we get states with populated domains favored by tensile (compressive) strains for Fig. \ref{fig:figure4}a (Fig. \ref{fig:figure4}c) after the external strain is removed at $\mu_0H=0\mathrm{~T}$. Then the external strain $\epsilon_{xx}$ is set to one of the target values as shown in the legend for Fig. \ref{fig:figure4}a and c, and a warming curve of $\rho_{xx}(T)$ is taken with the strain applied. After that, the $\Delta\rho_{xx}(T)$ curve under this specific external strain is obtained by subtracting the zero-strain background, and thus directly represents the temperature evolution of the nematic order parameter under the bias strain.

\section{Author Contributions}
T.L. performed elastoresistivity measurements and strain experiments. Z.F. grew the single crystals and conducted initial characterization.  T.L. carried out free-energy modeling and numerical simulation with M.Y. and L.Y.. The experiments were designed, and the results were analyzed through discussions among T.L., T.K., and L.Y.. T.L., T.K., and L.Y. wrote the manuscript with input from all authors.

\section{Data availability}
The datasets generated during and/or analyzed during the current study will be available in the Caltech Research Data Repository when the manuscript is published.
\section{Competing interests}
The authors declare no competing interests.

\bibliography{CoTa3S6}

@article{park2023tetrahedral,
  title={Tetrahedral triple-Q magnetic ordering and large spontaneous Hall conductivity in the metallic triangular antiferromagnet \Ce{Co1/3TaS2}},
  author={Park, Pyeongjae and Cho, Woonghee and Kim, Chaebin and An, Yeochan and Kang, Yoon-Gu and Avdeev, Maxim and Sibille, Romain and Iida, Kazuki and Kajimoto, Ryoichi and Lee, Ki Hoon and others},
  journal={Nat. Commun.},
  volume={14},
  number={1},
  pages={8346},
  year={2023},
  publisher={Nature Publishing Group UK London}
}

@article{takagi2023spontaneous,
  title   = {Spontaneous topological Hall effect induced by non-coplanar antiferromagnetic order in intercalated van der Waals materials},
  author  = {Takagi, H. and Takagi, R. and Minami, S. and Nomoto, T. and Ohishi, K. and Suzuki, M.-T. and Yanagi, Y. and Hirayama, M. and Khanh, N. D. and Karube, K. and Tokura, Y.},
  journal = {Nat. Phys.},
  volume  = {19},
  pages   = {961--968},
  year    = {2023}
}

@article{liu2024absence,
  title={Absence of ${E}_{2g}$ nematic instability and dominant ${A}_{1g}$ response in the kagome metal \ce{CsV3Sb5}},
  author={Liu, Zhaoyu and Shi, Yue and Jiang, Qianni and Rosenberg, Elliott W and DeStefano, Jonathan M and Liu, Jinjin and Hu, Chaowei and Zhao, Yuzhou and Wang, Zhiwei and Yao, Yugui and others},
  journal={Phys. Rev. X},
  volume={14},
  number={3},
  pages={031015},
  year={2024},
  publisher={APS}
}

@article{bohmer2022nematicity,
  title={Nematicity and nematic fluctuations in iron-based superconductors},
  author={B{\"o}hmer, Anna E and Chu, Jiun-Haw and Lederer, Samuel and Yi, Ming},
  journal={Nature Physics},
  volume={18},
  number={12},
  pages={1412--1419},
  year={2022},
  publisher={Nature Publishing Group UK London}
}

@article{little2020three,
  title={Three-state nematicity in the triangular lattice antiferromagnet \ce{Fe1/3NbS2}},
  author={Little, Arielle and Lee, Changmin and John, Caolan and Doyle, Spencer and Maniv, Eran and Nair, Nityan L and Chen, Wenqin and Rees, Dylan and Venderbos, J{\"o}rn WF and Fernandes, Rafael M and others},
  journal={Nat. Mater.},
  volume={19},
  number={10},
  pages={1062--1067},
  year={2020},
  publisher={Nature Publishing Group UK London}
}

@article{fradkin2010nematic,
  title={Nematic Fermi fluids in condensed matter physics},
  author={Fradkin, Eduardo and Kivelson, Steven A and Lawler, Michael J and Eisenstein, James P and Mackenzie, Andrew P},
  journal={Annu. Rev. Condens. Matter Phys.},
  volume={1},
  number={1},
  pages={153--178},
  year={2010},
  publisher={Annual Reviews}
}

@article{hwangbo2024strain,
  title   = {Strain tuning of vestigial three-state Potts nematicity in a correlated antiferromagnet},
  author  = {Hwangbo, Kyle and Rosenberg, Elliott and Cenker, John and Jiang, Qianni and Wen, Haidan and Xiao, Di and Chu, Jiun-Haw and Xu, Xiaodong},
  journal = {Nat. Phys.},
  year    = {2024},
  volume  = {20},
  pages   = {1888--1895},
  month   = oct
}

@article{montgomery1971method,
  title={Method for measuring electrical resistivity of anisotropic materials},
  author={Montgomery, HC},
  journal={J. Appl. Phys.},
  volume={42},
  number={7},
  pages={2971--2975},
  year={1971},
  publisher={AIP Publishing}
}

@article{park2022field,
  title   = {Field-tunable toroidal moment and anomalous Hall effect in noncollinear antiferromagnetic Weyl semimetal {Co$_{1/3}$TaS$_2$}},
  author  = {Park, Pyeongjae and Kang, Yoon-Gu and Kim, Junghyun and Lee, Ki Hoon and Noh, Han-Jin and Han, Myung Joon and Park, Je-Geun},
  journal = {npj Quantum Materials},
  volume  = {7},
  pages   = {42},
  year    = {2022}
}

@article{kuo2016ubiquitous,
  title={Ubiquitous signatures of nematic quantum criticality in optimally doped Fe-based superconductors},
  author={Kuo, Hsueh-Hui and Chu, Jiun-Haw and Palmstrom, Johanna C and Kivelson, Steven A and Fisher, Ian R},
  journal={Science},
  volume={352},
  number={6288},
  pages={958--962},
  year={2016},
  publisher={American Association for the Advancement of Science}
}

@article{park2024dft+,
  title={{DFT+ DMFT} study of the magnetic susceptibility and the correlated electronic structure in transition-metal intercalated \ce{NbS2}},
  author={Park, Hyowon and Martin, Ivar},
  journal={Phys. Rev.  B},
  volume={109},
  number={8},
  pages={085110},
  year={2024},
  publisher={APS}
}

@article{yanagi2023generation,
  title={Generation of modulated magnetic structures based on cluster multipole expansion: Application to $\alpha$-\ce{Mn} and \ce{CoM3S6}},
  author={Yanagi, Yuki and Kusunose, Hiroaki and Nomoto, Takuya and Arita, Ryotaro and Suzuki, Michi-To},
  journal={Phys. Rev.  B},
  volume={107},
  number={1},
  pages={014407},
  year={2023},
  publisher={APS}
}

@article{heinonen2022magnetic,
  title={Magnetic ground states of a model for \ce{MNb3S6} (\ce{ M= Co, Fe, Ni})},
  author={Heinonen, O and Heinonen, RA and Park, H},
  journal={Phys. Rev.  Materials},
  volume={6},
  number={2},
  pages={024405},
  year={2022},
  publisher={APS}
}

@article{parkin19803A,
  title={$3d$ transition-metal intercalates of the niobium and tantalum dichalcogenides. {I.} Magnetic properties},
  author={Parkin, SSP and Friend, RH},
  journal={Philos. Mag. B},
  volume={41},
  number={1},
  pages={65--93},
  year={1980},
  publisher={Taylor \& Francis}
}

@ARTICLE{Borzi2007-li,
  title     = "Formation of a nematic fluid at high fields in \ce{Sr3Ru2O7}",
  author    = "Borzi, R A and Grigera, S A and Farrell, J and Perry, R S and
               Lister, S J S and Lee, S L and Tennant, D A and Maeno, Y and
               Mackenzie, A P",
  journal   = "Science",
  publisher = "American Association for the Advancement of Science (AAAS)",
  volume    =  315,
  number    =  5809,
  pages     = "214--217",
  month     =  jan,
  year      =  2007
}

@ARTICLE{Hinkov2008-dh,
  title     = "Electronic liquid crystal state in the high-temperature
               superconductor {YBa$_2$Cu$_3$O$_{6.45}$}",
  author    = "Hinkov, V and Haug, D and Fauqué, B and Bourges, P and Sidis, Y
               and Ivanov, A and Bernhard, C and Lin, C T and Keimer, B",
  journal   = "Science",
  publisher = "American Association for the Advancement of Science (AAAS)",
  volume    =  319,
  number    =  5863,
  pages     = "597--600",
  month     =  feb,
  year      =  2008
}

@ARTICLE{Chu2010-cl,
  title     = "In-plane resistivity anisotropy in an underdoped iron arsenide
               superconductor",
  author    = "Chu, Jiun-Haw and Analytis, James G and De Greve, Kristiaan and
               McMahon, Peter L and Islam, Zahirul and Yamamoto, Yoshihisa and
               Fisher, Ian R",
  journal   = "Science",
  publisher = "American Association for the Advancement of Science (AAAS)",
  volume    =  329,
  number    =  5993,
  pages     = "824--826",
  month     =  aug,
  year      =  2010
}

@ARTICLE{Ghimire2018-bn,
  title    = "Large anomalous Hall effect in the chiral-lattice antiferromagnet
              \ce{CoNb3S6}",
  author   = "Ghimire, Nirmal J and Botana, A S and Jiang, J S and Zhang, Junjie
              and Chen, Y-S and Mitchell, J F",
  journal  = "Nat. Commun.",
  volume   =  9,
  number   =  1,
  pages    =  3280,
  month    =  aug,
  year     =  2018
}

@misc{Park2024,
  title  = {Contrasting dynamical properties of single-{Q} and triple-{Q} magnetic orderings in a triangular lattice antiferromagnet},
  author = {Park, Pyeongjae and Cho, Woonghee and Kim, Chaebin and An, Yeochan and Iida, Kazuki and Kajimoto, Ryoichi and Matin, Sakib and Zhang, Shang-Shun and Batista, Cristian D. and Park, Je-Geun},
  year   = {2024},
  month  = oct,
  note   = {arXiv:2410.02180 [cond-mat.str-el]}
}

@ARTICLE{Chu2012-vj,
  title     = "Divergent nematic susceptibility in an iron arsenide
               superconductor",
  author    = "Chu, Jiun-Haw and Kuo, Hsueh-Hui and Analytis, James G and
               Fisher, Ian R",
  journal   = "Science",
  publisher = "science.org",
  volume    =  337,
  number    =  6095,
  pages     = "710--712",
  month     =  aug,
  year      =  2012
}

@article{dos2011procedure,
  title={Procedure for measuring electrical resistivity of anisotropic materials: A revision of the Montgomery method},
  author={Dos Santos, CAM and De Campos, A and Da Luz, MS and White, BD and Neumeier, JJ and De Lima, BS and Shigue, CY},
  journal={J. Appl. Phys.},
  volume={110},
  number={8},
  year={2011},
  publisher={AIP Publishing}
}

@article{PhysRevB.103.184408,
  title = {Interplay of sample composition and anomalous Hall effect in \ce{Co_xNbS2}},
  author = {Mangelsen, S. and Zimmer, P. and N\"ather, C. and Mankovsky, S. and Polesya, S. and Ebert, H. and Bensch, W.},
  journal = {Phys. Rev. B},
  volume = {103},
  issue = {18},
  pages = {184408},
  numpages = {13},
  year = {2021},
  month = {May},
  publisher = {American Physical Society}
}

@article{PhysRevResearch.2.023051,
  title = {Giant anomalous Hall effect in quasi-two-dimensional layered antiferromagnet \ce{Co_{1/3}NbS2}},
  author = {Tenasini, Giulia and Martino, Edoardo and Ubrig, Nicolas and Ghimire, Nirmal J. and Berger, Helmuth and Zaharko, Oksana and Wu, Fengcheng and Mitchell, J. F. and Martin, Ivar and Forr\'o, L\'aszl\'o and Morpurgo, Alberto F.},
  journal = {Phys. Rev. Res.},
  volume = {2},
  issue = {2},
  pages = {023051},
  numpages = {8},
  year = {2020},
  month = {Apr},
  publisher = {American Physical Society}
}

@article{tanaka2022large,
  title={Large anomalous Hall effect induced by weak ferromagnetism in the noncentrosymmetric antiferromagnet \ce{CoNb3S6}},
  author={Tanaka, Hiroaki and Okazaki, Shota and Kuroda, Kenta and Noguchi, Ryo and Arai, Yosuke and Minami, Susumu and Ideta, Shinichiro and Tanaka, Kiyohisa and Lu, Donghui and Hashimoto, Makoto and others},
  journal={Phys. Rev.  B},
  volume={105},
  number={12},
  pages={L121102},
  year={2022},
  publisher={APS}
}

@article{morosan2007sharp,
  title={Sharp switching of the magnetization in \ce{Fe1/4TaS2}},
  author={Morosan, Emilia and Zandbergen, HW and Li, Lu and Lee, Minhyea and Checkelsky, JG and Heinrich, Michael and Siegrist, Theo and Ong, N Phuan and Cava, RJ},
  journal={Phys. Rev.  B},
  volume={75},
  number={10},
  pages={104401},
  year={2007},
  publisher={APS}
}

@article{miyadai1983magnetic,
  title={Magnetic Properties of \ce{Cr1/3NbS2}},
  author={Miyadai, Tomonao and Kikuchi, Katsuya and Kondo, Hiromitsu and Sakka, Shuzo and Arai, Masatoshi and Ishikawa, Yoshikazu},
  journal={J. Phys. Soc. Jpn.},
  volume={52},
  number={4},
  pages={1394--1401},
  year={1983},
  publisher={The Physical Society of Japan}
}

@article{wu2022highly,
  title={Highly tunable magnetic phases in transition-metal dichalcogenide \ce{Fe_{1/3+$\delta$}NbS2}},
  author={Wu, Shan and Xu, Zhijun and Haley, Shannon C and Weber, Sophie F and Acharya, Arani and Maniv, Eran and Qiu, Yiming and Aczel, Adam A and Settineri, Nicholas S and Neaton, Jeffrey B and others},
  journal={Phys. Rev.  X},
  volume={12},
  number={2},
  pages={021003},
  year={2022},
  publisher={APS}
}

@article{xie2022structure,
  title={Structure and magnetism of iron-and chromium-intercalated niobium and tantalum disulfides},
  author={Xie, Lilia S and Husremovic, Samra and Gonzalez, Oscar and Craig, Isaac M and Bediako, D Kwabena},
  journal={J. Am. Chem. Soc.},
  volume={144},
  number={22},
  pages={9525--9542},
  year={2022},
  publisher={ACS Publications}
}

@ARTICLE{Fernandes2019-kt,
  title     = "Intertwined vestigial order in quantum materials: Nematicity and
               beyond",
  author    = "Fernandes, Rafael M and Orth, Peter P and Schmalian, Jörg",
  journal   = "Annu. Rev. Condens. Matter Phys.",
  publisher = "Annual Reviews",
  volume    =  10,
  number    =  1,
  pages     = "133--154",
  month     =  mar,
  year      =  2019
}

@ARTICLE{Eckberg2020-so,
  title     = "Sixfold enhancement of superconductivity in a tunable electronic
               nematic system",
  author    = "Eckberg, Chris and Campbell, Daniel J and Metz, Tristin and
               Collini, John and Hodovanets, Halyna and Drye, Tyler and Zavalij,
               Peter and Christensen, Morten H and Fernandes, Rafael M and Lee,
               Sangjun and Abbamonte, Peter and Lynn, Jeffrey W and Paglione,
               Johnpierre",
  journal   = "Nat. Phys.",
  publisher = "Springer Science and Business Media LLC",
  volume    =  16,
  number    =  3,
  pages     = "346--350",
  year      =  2020
}

@ARTICLE{Frachet2022-bt,
  title     = "Elastoresistivity in the incommensurate charge density wave phase
               of \ce{BaNi2(As_{1-x}P_x)_2}",
  author    = "Frachet, M and Wiecki, P and Lacmann, T and Souliou, S M and
               Willa, K and Meingast, C and Merz, M and Haghighirad, A-A and Le
               Tacon, M and Böhmer, A E",
  journal   = "Npj Quantum Mater.",
  publisher = "Springer Science and Business Media LLC",
  volume    =  7,
  number    =  1,
  pages     = "1--8",
  month     =  dec,
  year      =  2022
}

@article{Ye2023,
  author = {Ye, Linda and Sun, Yue and Sunko, Veronika and Rodriguez-Nieva, Joaquin F. and Ikeda, Matthias S. and Worasaran, Thanapat and Sorensen, Matthew E. and Bachmann, Maja D. and Orenstein, Joseph and Fisher, Ian R.},
  title = {Elastocaloric signatures of symmetric and antisymmetric strain-tuning of quadrupolar and magnetic phases in DyB2C2},
  journal = {Proc. Natl. Acad. Sci. U.S.A.},
  volume = {120},
  pages = {e2302800120},
  year = {2023}
}

@article{Straquadine2022,
  author = {Straquadine, J. A. W. and Ikeda, M. S. and Fisher, I. R.},
  title = {Evidence for realignment of the charge density wave state in ${\mathrm{ErTe}}_{3}$ and ${\mathrm{TmTe}}_{3}$ under uniaxial stress via elastocaloric and elastoresistivity measurements},
  journal = {Phys. Rev. X},
  volume = {12},
  pages = {021046},
  year = {2022}
}

@ARTICLE{Fernandes2020-us,
  title     = "Nematicity with a twist: Rotational symmetry breaking in a moiré
               superlattice",
  author    = "Fernandes, Rafael M and Venderbos, Jörn W F",
  journal   = "Sci. Adv.",
  publisher = "American Association for the Advancement of Science (AAAS)",
  volume    =  6,
  number    =  32,
  pages     = "eaba8834",
  month     =  aug,
  year      =  2020
}

@ARTICLE{Cao2021-sv,
  title     = "Nematicity and competing orders in superconducting magic-angle
               graphene",
  author    = "Cao, Yuan and Rodan-Legrain, Daniel and Park, Jeong Min and Yuan,
               Noah F Q and Watanabe, Kenji and Taniguchi, Takashi and
               Fernandes, Rafael M and Fu, Liang and Jarillo-Herrero, Pablo",
  journal   = "Science",
  publisher = "American Association for the Advancement of Science (AAAS)",
  volume    =  372,
  number    =  6539,
  pages     = "264--271",
  month     =  apr,
  year      =  2021
}

@ARTICLE{Xu2022-hu,
  title     = "Three-state nematicity and magneto-optical Kerr effect in the
               charge density waves in kagome superconductors",
  author    = "Xu, Yishuai and Ni, Zhuoliang and Liu, Yizhou and Ortiz, Brenden
               R and Deng, Qinwen and Wilson, Stephen D and Yan, Binghai and
               Balents, Leon and Wu, Liang",
  journal   = "Nat. Phys.",
  publisher = "Nature Publishing Group",
  volume    =  18,
  number    =  12,
  pages     = "1470--1475",
  month     =  nov,
  year      =  2022
}

@ARTICLE{Li2023-ag,
  title     = "Electronic nematicity without charge density waves in
               titanium-based kagome metal",
  author    = "Li, Hong and Cheng, Siyu and Ortiz, Brenden R and Tan, Hengxin
               and Werhahn, Dominik and Zeng, Keyu and Johrendt, Dirk and Yan,
               Binghai and Wang, Ziqiang and Wilson, Stephen D and Zeljkovic,
               Ilija",
  journal   = "Nat. Phys.",
  publisher = "Springer Science and Business Media LLC",
  volume    =  19,
  number    =  11,
  pages     = "1591--1598",
  month     =  nov,
  year      =  2023
}

@ARTICLE{Fernandes2014-qg,
  title     = "What drives nematic order in iron-based superconductors?",
  author    = "Fernandes, R M and Chubukov, A V and Schmalian, J",
  journal   = "Nat. Phys.",
  publisher = "Nature Publishing Group",
  volume    =  10,
  number    =  2,
  pages     = "97--104",
  month     =  jan,
  year      =  2014
}

@article{duStronglyAnisotropicTransport1999,
  title = {Strongly Anisotropic Transport in Higher Two-Dimensional Landau Levels},
  author = {Du, R. R. and Tsui, D. C. and Stormer, H. L. and Pfeiffer, L. N. and Baldwin, K. W. and West, K. W.},
  year = 1999,
  month = jan,
  journal = {Solid State Communications},
  volume = {109},
  number = {6},
  pages = {389--394},
  issn = {0038-1098},
  urldate = {2026-01-14}
}

@article{lillyEvidenceAnisotropicState1999,
  title = {Evidence for an Anisotropic State of Two-Dimensional Electrons in High Landau Levels},
  author = {Lilly, M. P. and Cooper, K. B. and Eisenstein, J. P. and Pfeiffer, L. N. and West, K. W.},
  year = 1999,
  month = jan,
  journal = {Physical Review Letters},
  volume = {82},
  number = {2},
  pages = {394--397},
  publisher = {American Physical Society},
  urldate = {2026-01-14}
}

@article{farhangDiscoveryIntermediate2025,
  title = {Discovery of an Intermediate Nematic State in a Bilayer Kagome Metal {ScV$_6$Sn$_6$}},
  author = {Farhang, Camron and Meier, William R. and Lu, Weihang and Li, Jiangxu and Wu, Yudong and Mozaffari, Shirin and Madhogaria, Richa P. and Zhang, Yang and Mandrus, David and Xia, Jing},
  year = 2025,
  month = aug,
  journal = {Nature Communications},
  volume = {16},
  number = {1},
  pages = {7867},
  publisher = {Nature Publishing Group},
  issn = {2041-1723},
  urldate = {2026-01-16},
  copyright = {2025 The Author(s)},
  langid = {english}
}

@article{bohmerNematicityNematic2022,
  title = {Nematicity and Nematic Fluctuations in Iron-Based Superconductors},
  author = {B{\"o}hmer, Anna E. and Chu, Jiun-Haw and Lederer, Samuel and Yi, Ming},
  year = 2022,
  month = dec,
  journal = {Nature Physics},
  volume = {18},
  number = {12},
  pages = {1412--1419},
  publisher = {Nature Publishing Group},
  issn = {1745-2481},
  urldate = {2026-01-16},
  copyright = {2022 Springer Nature Limited},
  langid = {english}
}

@article{tanObservationThreeStateNematicity2024,
  title = {Observation of Three-State Nematicity and Domain Evolution in Atomically Thin Antiferromagnetic {NiPS$_3$}},
  author = {Tan, Qishuo and Occhialini, Connor A. and Gao, Hongze and Li, Jiaruo and Kitadai, Hikari and Comin, Riccardo and Ling, Xi},
  year = 2024,
  month = jun,
  journal = {Nano Letters},
  volume = {24},
  number = {24},
  pages = {7166--7172},
  publisher = {American Chemical Society},
  issn = {1530-6984},
  urldate = {2025-12-16}
}

@misc{Yao2025Potts,
  title = {Signatures of Three-State Potts Nematicity in Spin Excitations of the van der Waals Antiferromagnet {FePSe$_3$}},
  author = {Yao, Weiliang and Antonio, Viviane Pe{\c c}anha and Adroja, Devashibhai and Alvarado, S. J. Gomez and Gao, Bin and Xu, Sijie and Liu, Ruixian and Lu, Xingye and Dai, Pengcheng},
  year = {2025},
  note = {arXiv:2509.02475}
}

@article{chenUnidirectionalSpin2018,
  title = {Unidirectional Spin Density Wave State in Metallic {(Sr$_{1-x}$La$_x$)$_2$IrO$_4$}},
  author = {Chen, Xiang and Schmehr, Julian L. and Islam, Zahirul and Porter, Zach and Zoghlin, Eli and Finkelstein, Kenneth and Ruff, Jacob P. C. and Wilson, Stephen D.},
  year = 2018,
  month = jan,
  journal = {Nature Communications},
  volume = {9},
  number = {1},
  pages = {103},
  publisher = {Nature Publishing Group},
  issn = {2041-1723},
  urldate = {2026-01-16},
  copyright = {2017 The Author(s)},
  langid = {english}
}

@article{guoSpectralEvidence2023,
 title = {Spectral Evidence for Unidirectional Charge Density Wave in Detwinned {BaNi$_2$As$_2$}},
  author = {Guo, Yucheng and Klemm, Mason and Oh, Ji Seop and Xie, Yaofeng and Lei, Bing-Hua and Moreschini, Luca and Chen, Cheng and Yue, Ziqin and Gorovikov, Sergey and Pedersen, Tor and Michiardi, Matteo and Zhdanovich, Sergey and Damascelli, Andrea and Denlinger, Jonathan and Hashimoto, Makoto and Lu, Donghui and Jozwiak, Chris and Bostwick, Aaron and Rotenberg, Eli and Mo, Sung-Kwan and Moore, Rob G. and Kono, Junichiro and Birgeneau, Robert J. and Singh, David J. and Dai, Pengcheng and Yi, Ming},
  year = 2023,
  month = aug,
  journal = {Physical Review B},
  volume = {108},
  number = {8},
  pages = {L081104},
  publisher = {American Physical Society},
  urldate = {2026-01-16}
}

@article{kuoMeasurementElastoresistivity2013,
  title = {Measurement of the Elastoresistivity Coefficients of the Underdoped Iron Arsenide {Ba(Fe$_{0.975}$Co$_{0.025}$)$_2$As$_2$}},
  author = {Kuo, Hsueh-Hui and Shapiro, Maxwell C. and Riggs, Scott C. and Fisher, Ian R.},
  year = 2013,
  month = aug,
  journal = {Physical Review B},
  volume = {88},
  number = {8},
  pages = {085113},
  publisher = {American Physical Society},
  urldate = {2026-01-16}
}

@article{rosenbergDivergenceQuadrupolestrain2019,
  title = {Divergence of the Quadrupole-Strain Susceptibility of the Electronic Nematic System {YbRu$_2$Ge$_2$}},
  author = {Rosenberg, Elliott W. and Chu, Jiun-Haw and Ruff, Jacob P. C. and Hristov, Alexander T. and Fisher, Ian R.},
  year = 2019,
  month = apr,
  journal = {Proceedings of the National Academy of Sciences},
  volume = {116},
  number = {15},
  pages = {7232--7237},
  publisher = {Proceedings of the National Academy of Sciences},
  urldate = {2026-01-16}
}

@misc{fengNonvolatileNematic2025,
  title  = {Nonvolatile Nematic Order Manipulated by Strain and Magnetic Field in a Layered Antiferromagnet},
  author = {Feng, Zili and Lu, Weihang and Lu, Tao and Liu, Fangyan and Sheeran, Joseph R. and Ye, Mengxing and Xia, Jing and Kurumaji, Takashi and Ye, Linda},
  year   = {2025},
  month  = jul,
  note   = {arXiv:2507.05486 [cond-mat.str-el]}
}

@article{riggsEvidenceNematic2015,
  title = {Evidence for a Nematic Component to the Hidden-Order Parameter in {URu$_2$Si$_2$} from Differential Elastoresistance Measurements},
  author = {Riggs, Scott C. and Shapiro, M. C. and Maharaj, Akash V. and Raghu, S. and Bauer, E. D. and Baumbach, R. E. and {Giraldo-Gallo}, P. and Wartenbe, Mark and Fisher, I. R.},
  year = 2015,
  month = mar,
  journal = {Nature Communications},
  volume = {6},
  number = {1},
  pages = {6425},
  publisher = {Nature Publishing Group},
  issn = {2041-1723},
  urldate = {2026-01-20},
  copyright = {2015 Springer Nature Limited},
  langid = {english}
}

@book{luthiPhysicalAcoustics2007,
  title = {Physical Acoustics in the Solid State},
  author = {L{\"u}thi, Bruno},
  year = 2007,
  month = aug,
  publisher = {Springer Science \& Business Media},
  isbn = {978-3-540-72194-9},
  langid = {english}
}

@misc{sunIntertwinedChargeDensity2025,
  title = {Intertwined Charge Density Wave Order in Vanadium Based Kagome Superconductors},
  author = {Sun, Kuanglv and Nie, Linpeng and Li, Hongyu and Zhao, Jiyin and Rao, Huachen and Yu, Fanghang and Shi, Mengzhu and Xiang, Ziji and Ying, Jianjun and Wang, Zhenyu and Wu, Tao and Chen, Xianhui},
  year = 2025,
  month = nov,
  publisher = {Research Square},
  issn = {2693-5015},
  urldate = {2026-01-03},
  archiveprefix = {Research Square}
}

@article{
doi:10.1073/pnas.2105911118,
author = {Matthias S. Ikeda  and Thanapat Worasaran  and Elliott W. Rosenberg  and Johanna C. Palmstrom  and Steven A. Kivelson  and Ian R. Fisher },
title = {Elastocaloric signature of nematic fluctuations},
journal = {Proceedings of the National Academy of Sciences},
volume = {118},
number = {37},
pages = {e2105911118},
year = {2021}}

\end{document}